\documentclass{elsart}
\usepackage{epsfig}
\usepackage{enumerate}

\begin{document}
\begin{frontmatter}

\title{Noise-induced absorbing phase transition in a model of opinion formation}

\author{Allan R. Vieira}
\thanks{allanrv@if.uff.br}
\author{and Nuno Crokidakis}
\thanks{nuno@if.uff.br}

\address{
Instituto de F\'{\i}sica, \hspace{1mm} Universidade Federal Fluminense \\
Av. Litor\^anea s/n, \hspace{1mm} 24210-340 \hspace{1mm} Niter\'oi - RJ, \hspace{1mm} Brazil}

\maketitle

\begin{abstract}
\noindent
In this work we study a 3-state ($+1$, $-1$, $0$) opinion model in the presence of noise and disorder. We consider pairwise competitive interactions, with a fraction $p$ of those interactions being negative (disorder). Moreover, there is a noise $q$ that represents the probability of an individual spontaneously change his opinion to the neutral state. Our aim is to study how the increase/decrease of the fraction of neutral agents affects the critical behavior of the system and the evolution of opinions. We derive analytical expressions for the order parameter of the model, as well as for the stationary fraction of each opinion, and we show that there are distinct phase transitions. One is the usual ferro-paramagnetic transition, that is in the Ising universality class. In addition, there are para-absorbing and ferro-absorbing transitions, presenting the directed percolation universality class. Our results are complemented by numerical simulations.

\end{abstract}
\end{frontmatter}


\section{Introduction}

\qquad The study of dynamics of opinion formation is nowadays a hot topic in the Statistical Physics of Complex Systems, with a considerable amount of papers published in the last years (see \cite{galam_book,sen_book,pmco_book,rmp} and references therein). Even simple models can exhibit an interesting collective behavior that emerges from the microscopic interaction among individuals or agents in a given social network. Usually those models exhibit nonequilibrium phase transitions and rich critical phenomena, which justifies the interest of physicists in the study of opinion dynamics \cite{galam_book,sen_book,pmco_book,rmp,nuno_jstat_2013,nuno_bjp,lccc,biswas2,sen,diao,wu}.

In the last few years, a recent attention has been done to the kinetic exchange opinion models (KEOM) \cite{lccc,biswas2,sen,biswas3}, inspired in models of wealth exchange \cite{chakra1,chakra2,chatt}. The LCCC model was the first one to consider kinetic exchanges among pairs of agents that present continuous states (opinions) \cite{lccc}. In this case, the model presents a continuous symmetry-breaking phase transition. After that, some extensions were analyzed for continuous and discrete opinions. For example, the inclusion of competitive interactions \cite{biswas2}, three-agents' interactions \cite{biswas3}, dynamic self-confidence \cite{xiong}, presence of inflexible agents \cite{nuno_celia_victor}, and others, similarly to was done previously in other opinion dynamics, like the Galam's models \cite{major-rule,galam_jacobs}. In all these extensions the critical behavior of the system was extensively analyzed.

Dynamics of decision-making has been treated in several works in Psychology \cite{psyco1,psyco2} and Neuroscience \cite{neuro1,neuro2,neuro3}. For the dynamics of opinion formation, we find many models by physicists dedicated to explain the decision-making process or the exchange of opinion through interactions among agents \cite{rmp}. The mechanisms consider kinetic exchanges (KEOM \cite{lccc,biswas2,sen,biswas3}), imitation (voter model \cite{voter}, Sznajd model \cite{sznajd}) or the power of local majorities (majority-rule model \cite{major-rule}, majority-vote model \cite{major-voter}), among others. Nevertheless, the inclusion of noise and disorder can be considered in such models \cite{galam_book,sen_book,pmco_book,rmp}.

Usually discrete opinion models consider two distinct positions or opinions $o=\pm 1$ (yes or no, democrat or republican, candidate A or candidate B). They can be enriched with the inclusion of a third state, $o=0$, representing neutral state or indecision. Indecision is a current and rising phenomenon which affects both recent and consolidated democracies \cite{indecision1}. Many reasons can lead an individual to become neutral or undecided, for example it can be associated to an anticonformism/nonconformism to the proposals on both sides of the debate. The impact of indecision/neutrality was considered recently in many works \cite{biswas2,biswas3,nuno_celia_victor,indecision1,indecision2,indecision3,indecision4,indecision5,indecision6}.

In this work we consider a discrete KEOM in the presence of noise and disorder. In addition to pairwise random interactions, we introduce an indecision noise that significantly affects the dynamics of the system. Our aim is to analyze the critical behavior of the model. In this case, based on analytical and numerical results, we found three distinct phase transitions, namely the usual ferro-paramagnetic transition, and two distinct transitions to an absorbing state: from the ferromagnetic state and from the paramagnetic one.


\section{Model and Results}

\qquad We considered a KEOM \cite{lccc,biswas2,biswas3} with competitive positive/negative interactions. Our artificial society is represented by $N$ individuals in a fully-connected graph. Each agent $i$ can be in one of three possible opinions at each time step $t$, i.e., $o_{i}(t)=+1, -1$ or $0$. This general scheme can represent a public debate with two distinct choices, for example \textit{yes} and \textit{no}, and also including the undecided/neutral state. The following microscopic rules control our model:
\begin{enumerate}
\item we choose two agents at random, say $i$ and $j$, in a way that $j$ will try to persuade $i$;
\item with probability $1-q$, the opinion of agent $i$ in the next step $t+1$ is updated according to the kinetic rule $o_{i}(t+1) = sgn[o_{i}(t) + \mu_{ij}o_{j}(t)]$;
\item with probability $q$, the agent $i$ spontaneously change to the neutral state, i.e., $o_{i}(t+1)=0$.
\end{enumerate}
In the above dynamic rule, sgn(x) is the signal function defined such that sgn(0)=0. This is usual in KEOM, in order to keep all the agents' opinions in one the three possible ones, $+1$, $-1$ or $0$ \cite{biswas2,biswas3,nuno_celia_victor}. The pairwise couplings $\mu_{ij}$ are quenched random variables \footnote{The nature of the random variables $\mu_{ij}$ does not affect our results, they can also be considered as annealed variables.} that follows the discrete probability distribution $F(\mu_{ij})=p\,\delta(\mu_{ij}+1)+(1-p)\,\delta(\mu_{ij}-1)$. In other words, the parameter $p$ stands for the fraction of negative interactions. As discussed in previous works \cite{biswas2,nuno_celia_victor}, the consideration of such negative interactions produces an effect similiar to the introduction of Galam's contrarians in the population \cite{major-rule,contrarians}. In addition, competitive interactions were also considered for the modelling of coalition forming \cite{galam_coal}. The probability $q$ acts as a noise in the system, and it allows an autonomous decision of an individual to become neutral \cite{nuno_jstat_2013,ben-naim}. It can be viewed as the volatility of some individuals, who tend to spontaneously change their choices. In a two-candidate election, if a given individual does not agree with the arguments of supporters of both sides, he/she can decide to not vote for any candidate, and in this case he/she becomes neutral. In this case, this indecision noise must be differentiated/disassociated from other usual kinds of noises because, unlike the others, it privileges only the neutral opinion. As a recent example, in the 2012 USA election Barack Obama and Mitt Romney disputed for the election for president as the main candidates. It was reported that two months out from election day, nearly a quarter of all registered voters are either undecided about the presidential race or iffy in their support for a candidate, as indicated by polls \cite{fox}.

For $q=0$, i.e., in the absence of noise, the model undergoes a nonequilibrium order-disorder (or ferro-paramagnetic) transition at a critical fraction $p_{c}=1/4$ \cite{biswas2}. In the ordered ferromagnetic phase, one of the extreme opinions $+1$ or $-1$ dominates the population, whereas in the disordered paramagnetic phase the three opinions coexist with equal fractions ($1/3$). 

At this point, some definitions are necessary. The order parameter of the system can be defined as
\begin{equation} \label{eq1}
O = \left \langle\frac{1}{N}\left|\sum_{i=1}^{N} o_{i}\right|\right \rangle ~,
\end{equation}
\noindent
that is the ``magnetization per spin'' of the system, and $\langle\, ...\, \rangle$ stands for average over disorder or configurations, computed at the steady states. Let us also define $f_{1},f_{-1}$ and $f_{0}$ as the stationary fractions or densities of opinions $+1, -1$ and $0$, respectively. 

One can start considering the probabilities that contribute to increase and decrease the order parameter. Following \cite{biswas2,biswas3}, one can obtain the master equation for $O$, 
\begin{eqnarray}\label{eq2} \nonumber
\frac{d}{dt}\,O & = & qf_{-1} + (1-q)[ (1-p)f_1 f_{-1} +pf_{-1}^{2} + (1-p)f_0 f_1 +pf_0 f_{-1}] - \\ \nonumber
&   &\mbox{}- qf_1  -(1-q) [(1-p)f_1 f_{-1} + pf_{1}^{2} + (1-p)f_0 f_{-1} + \\
&   &\mbox{}+ pf_0 f_{1}] = 0 ~.
\end{eqnarray}
In the stationary state $dO/dt=0$. Using the normalization condition $f_{1} + f_{-1} + f_{0} = 1$, we obtain two solutions for Eq. (\ref{eq2}) in the stationary state, namely $2f_{1}+f_0=1$, which implies in $f_{1}=f_{-1}=(1-f_{0})/2$ (disordered solution), or 
\begin{equation}\label{eq3}
f_{0} =  \frac{q+p(1-q)}{(1-p)(1-q)}.
\end{equation}

In this case, Eq. (\ref{eq3}) is valid in the ferromagnetic phase. We emphasize that $q=0$ leads to $f_{0} = p/(1-p)$, which agrees with the result of Ref. \cite{biswas2}. One can obtain another equation for $f_{0}$ considering the fluxes into and out of the neutral state $o=0$. In this case, the master equation for $f_{0}$ is given by
\begin{eqnarray}  \label{eq4} \nonumber
\frac{d}{dt}\,f_{0} & = & q(f_{1}+f_{-1})+p(1-q)(f_{1}^{2}+f_{-1}^{2})+2(1-p)(1-q)f_{1}f_{-1} - \\
&   &\mbox{} - (1-p)(1-q)f_{0}(f_{1}+f_{-1}) - p(1-q)f_{0}(f_{1}+f_{-1})~.
\end{eqnarray}
Considering the disordered phase, where $f_{1}=f_{-1}=(1-f_{0})/2$, Eq. (\ref{eq4}) gives us in the stationary state (where $df_{0}/dt=0$)
\begin{equation}  \label{eq5}
(1-q)\left(\frac{1-f_{0}}{2}\right)^{2} =[(1-q)f_{0}-q]\left(\frac{1-f_{0}}{2}\right) ~,
\end{equation}
\noindent
which gives us two solutions, namely $f_{0}=1$ which can be ignored by considering the steady state of the other two fractions $f_{1}$ and $f_{-1}$ \cite{biswas2}, or
\begin{equation}  \label{eq6}
f_{0} = \frac{1+q}{3(1-q)}.
\end{equation}
In this case, Eq. (\ref{eq6}) is valid in the paramagnetic phase. The above equations (\ref{eq3}) and (\ref{eq6}) are both valid at the critical point, and we can equate them to obtain
\begin{equation} \label{eq7}
q_{c}(p) = \frac{1-4p}{2(1-p)} ~.
\end{equation}
\noindent
These critical noises separate the ferromagnetic and the paramagnetic phases. As discussed above, in the ferromagnetic phase one of the extreme opinions $+1$ or $-1$ dominates the population (one of the sides wins the debate), whereas in the paramagnetic phase the two extreme opinions coexist ($f_{1}=f_{-1}$, i.e., there is no decision). Notice that we recover $f_{0}=1/3$ in Eq. (\ref{eq6}) and $p_{c}=1/4$ in Eq. (\ref{eq7}) for $q=0$, in agreement with \cite{biswas2}.

In order to obtain an analytical expression for the order parameter, one can consider the fluxes into and out of the state $o=+1$. The master equation for $f_{1}$ is then
\begin{eqnarray}  \label{eq8} \nonumber
\frac{d}{dt}\,f_{1} & = & (1-q)\left[(1-p)f_{0}f_{1}  + pf_{0}f_{-1}\right] - (1-q)\left[pf_{1}^{2} + (1-p)f_{1} f_{-1}\right] - \\ 
&   &\mbox{}-qf_{1} ~.
\end{eqnarray}
Considering the normalization condition and the expression for $f_{0}$ valid in the ferromagnetic phase, Eq. (\ref{eq3}), we obtain for $f_{1}$ in the stationary state (where $df_{1}/dt=0$)
\begin{equation}  \label{eq9}
f_{1} =\frac{(2p-1)[1-2(p+q-pq)] \pm \sqrt{\Delta}}{2(1-p)(2p-1)} ~,
\end{equation}
\noindent
where 
\begin{equation}  \label{eq10}
\Delta=(1-2p)[1-2(p+q-pq)][1-2(2p+q-pq)] ~.
\end{equation}

\begin{figure}[t]
\begin{center}
\vspace{6mm}
\includegraphics[scale=0.33,angle=-90]{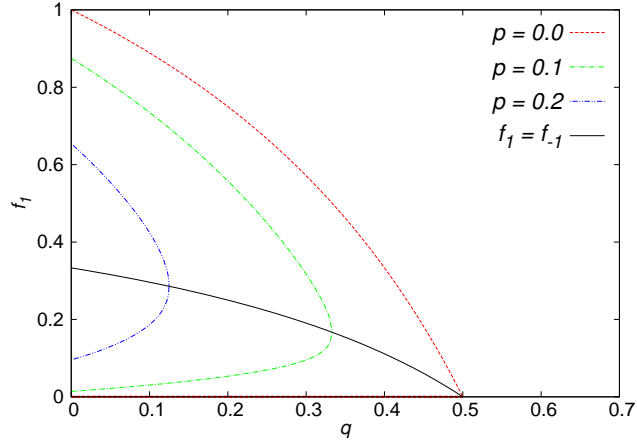}
\end{center}
\caption{(Color online) The stationary fraction $f_{1}$ of opinions $o=+1$ versus $q$ for typical values of $p$, obtained from Eq. (\ref{eq9}). For a fixed $p$, if we choose $q<q_c(p)$ we have two possible solutions for $f_{1}$, as discussed in the text. The solid line $f_{1}=f_{-1}$ is the solution for $f_{1}$ in the paramagnetic phase, given by Eq. (\ref{eq16}).}
\label{fig1}
\end{figure}

Eq. (\ref{eq9}) is plotted in Fig. \ref{fig1} as a function of $q$ for typical values of $p$. As Eq. (\ref{eq9}) predicts two solutions (see the $\pm$ signals), one has two curves for each value of $p$ since $q<q_{c}(p)$, where $q_{c}(p)$ is given by Eq. (\ref{eq7}). When $f_1$ assumes one of these values consequently $f_{-1}$ takes the other one. The curve labeled as $f_{1}=f_{-1}$ is the disordered paramagnetic solution for the stationary fractions $f_{1}$ and $f_{-1}$. This solution signals the limit of validity of Eq. (\ref{eq9}), and it will be discussed in the following.

The order parameter can be given by $O=|f_{1}-f_{-1}|=|2f_{1}+f_{0}-1|$. Considering Eqs. (\ref{eq3}) and (\ref{eq9}) for $f_{0}$ and $f_{1}$, respectively, one obtains 
\begin{equation} \label{eq11}
O = \frac{\sqrt{\left [ 1-2(p+q-pq) \right ] \left [ 1-2(2p+q-pq)\right ] }}{(1-p)(1-q) \sqrt{1-2p} } ~.
\end{equation}
\noindent
One can see from Eq. (\ref{eq11}) that consensus is reached only for $p=q=0$, i.e., in the absence of negative interactions and noise all agents of the system will share one of the extreme opinions, $+1$ or $-1$. In order to obtain the critical exponent $\beta$ that governs the order parameter in the vicinity of the order-disorder phase transition, one can simplify Eq. (\ref{eq11}) using Eqs. (\ref{eq3}) and (\ref{eq7}). In this case, we have
\begin{eqnarray}  \label{eq13}
O = \sqrt{\frac{2(1-f_0)(q-q_c)}{(1-q)(2p-1)}} ~, 
\end{eqnarray}
\noindent
where $f_{0}$ is given by Eq. (\ref{eq3}), i.e., the solution valid in the ferromagnetic phase. In other words, one can write the order parameter in the usual form $O \sim (q-q_c)^{\beta}$, where $\beta = 1/2$, a typical Ising mean-field exponent in a ferro-paramagnetic phase transition, suggesting that our model is in the same universality class of the mean-field Ising model, as expected due to the mean-field character of the interactions.

On the other hand, the case $p=0$ presents a distinct behavior. Putting $p=0$ in Eq. (\ref{eq3}), one obtains
\begin{eqnarray}  \label{eq14}
f_{0}(p=0) & = & \frac{q}{1-q} ~.
\end{eqnarray}
Using this result, $q_{c}(p=0)=1/2$ obtained from Eq. (\ref{eq7}) and putting $p=0$ in Eq. (\ref{eq13}), the order parameter can be written as 
\begin{eqnarray}  \label{eq15}
O(p=0) & = & \frac{2q - 1}{q-1} ~,
\end{eqnarray}
and thus $O(p=0) \sim (q-q_{c})^{\beta}$, where $\beta=1$ and $q_c=1/2$. Furthermore, using $q=q_{c}=1/2$ in Eq. (\ref{eq14}), we have $f_{0}=1$, which implies in $f_{1}=f_{-1}=0$ due to the normalization condition. This result implies that for $q>1/2$ and $p=0$ all agents will be in the neutral state $o=0$. A look to the microscopic rules that define our model shows that the system will remain in this state forever, indicating an absorbing state. Thus, we have for $p=0$ an active-absorbing (ferromagnetic-absorbing) phase transition, i.e., the critical behavior is affected and the system can be mapped in the universality class of mean-field directed percolation (or contact process) \cite{absorbing_book,dickman,hinrichsen}. 

As previous discussed, if $p\neq 0$ the order parameter goes to zero with $\beta = 1/2$, signaling a ferromagnetic-paramagnetic phase transition at critical points $q_{c}(p)$ given by Eq. (\ref{eq7}). For $q>q_{c}(p)$ the ferromagnetic solution for $f_0$, Eq. (\ref{eq3}) is not valid anymore. In this case the valid solution for $f_{0}$ is given by Eq. (\ref{eq6}), the solution in the paramagnetic phase. Considering Eq. (\ref{eq6}) and that in the paramagnetic phase we have $f_{1}=f_{-1}=(1-f_{0})/2$, one obtains 
\begin{equation} \label{eq16}
f_{1}=f_{-1}=\frac{1-2q}{3(1-q)} ~,
\end{equation}

\begin{figure}[t]
\begin{center}
\vspace{6mm}
\includegraphics[scale=0.33,angle=-90]{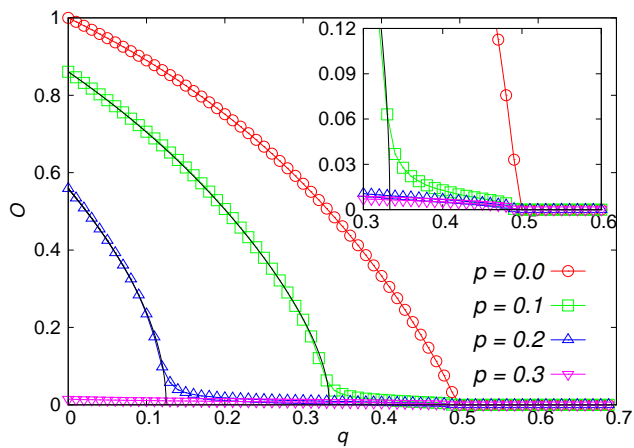}
\end{center}
\caption{(Color online) Order parameter $O$ versus noise $q$ for typical values of the fraction $p$ of negative interactions. The symbols are results of numerical simulations, and the lines are given by Eq. (\ref{eq11}). For $p<p_{c}=1/4$ the system undergoes a phase transition at critical points given by Eq. (\ref{eq7}). For $p>0$ the model presents a ferro-paramagnetic transition at points given by Eq. (\ref{eq7}) and a paramagnetic-absorbing transition at $q_{c}=1/2$. The inset shows that we have $O=0$ for $q\geq 1/2$ for all values of $p$. The population size is $N=10^{4}$, and results are averaged over $100$ independent simulations.}
\label{fig2}
\end{figure}

\noindent
that is the above-mentioned solution for $f_{1}=f_{-1}$ in the paramagnetic phase, see Fig. \ref{fig1}. From Eq. (\ref{eq16}), one can see that $f_{1}<0$ and $f_{-1}<0$ for $q> 1/2$. These solutions are not physically acceptable, and thus the valid solution for $q> 1/2$ is $f_{1}=f_{-1}=0$ and $f_{0}=1$. In addition, as $O=0$ in the paramagnetic phase, we can use another order parameter to analyze the system in the vicinity of $q=1/2$, namely
\begin{equation} \label{eq17}
\bar{O} =  \frac{1-f_0}{2} = \frac{1-2q}{3(1-q)} ~,  
\end{equation}
where we used Eq. (\ref{eq6}). In this case, we have $\bar{O}=0$ in the absorbing phase (where $f_{0}=1$) and $\bar{O}>0$ in the paramagnetic phase (where $f_{0}<1$). One can rewrite Eq. (\ref{eq17}) as $\bar{O} \sim (q-\bar{q}_{c})^{\beta} $ where $\beta = 1$ and $\bar{q}_c = 1/2$. Thus, for $q>\bar{q}_{c}=1/2$ the system is always in an absorbing phase with all individuals sharing the neutral state $o=0$ (or in other words $f_{0}=1$), independent of $p$. Thus, the transition may be of active-absorbing type for $p=0$, or paramagnetic-absorbing for $p>0$, both occurring at $q_{c}=1/2$ and belonging to the directed percolation universality class (critical exponent $\beta=1$) \cite{absorbing_book,dickman,hinrichsen}. For the best of our knowledge, it is the first time that the para-absorbing transition appears in a KEOM.

To complement our results, we performed numerical simulations for a population size $N=10^{4}$. We computed the order parameter $O$ by Eq. (\ref{eq1}). In Fig. \ref{fig2} we exhibit the numerical results for $O$ versus $q$ and typical values of $p$, together with the analytical result given by Eq. (\ref{eq11}). One can observe transitions at different points $q_{c}$ that depend on $p$, with the usual finite-size effects for $p=0.1$, $0.2$ and $0.3$ (see the inset). Moreover, we can also see in the inset that the order parameter goes exactly to zero for $q\geq 1/2$, independent of the value of $p$, confirming the transition to the absorbing state, as analytically predicted. 

\begin{figure}[t]
\begin{center}
\vspace{6mm}
\includegraphics[scale=0.25,angle=-90]{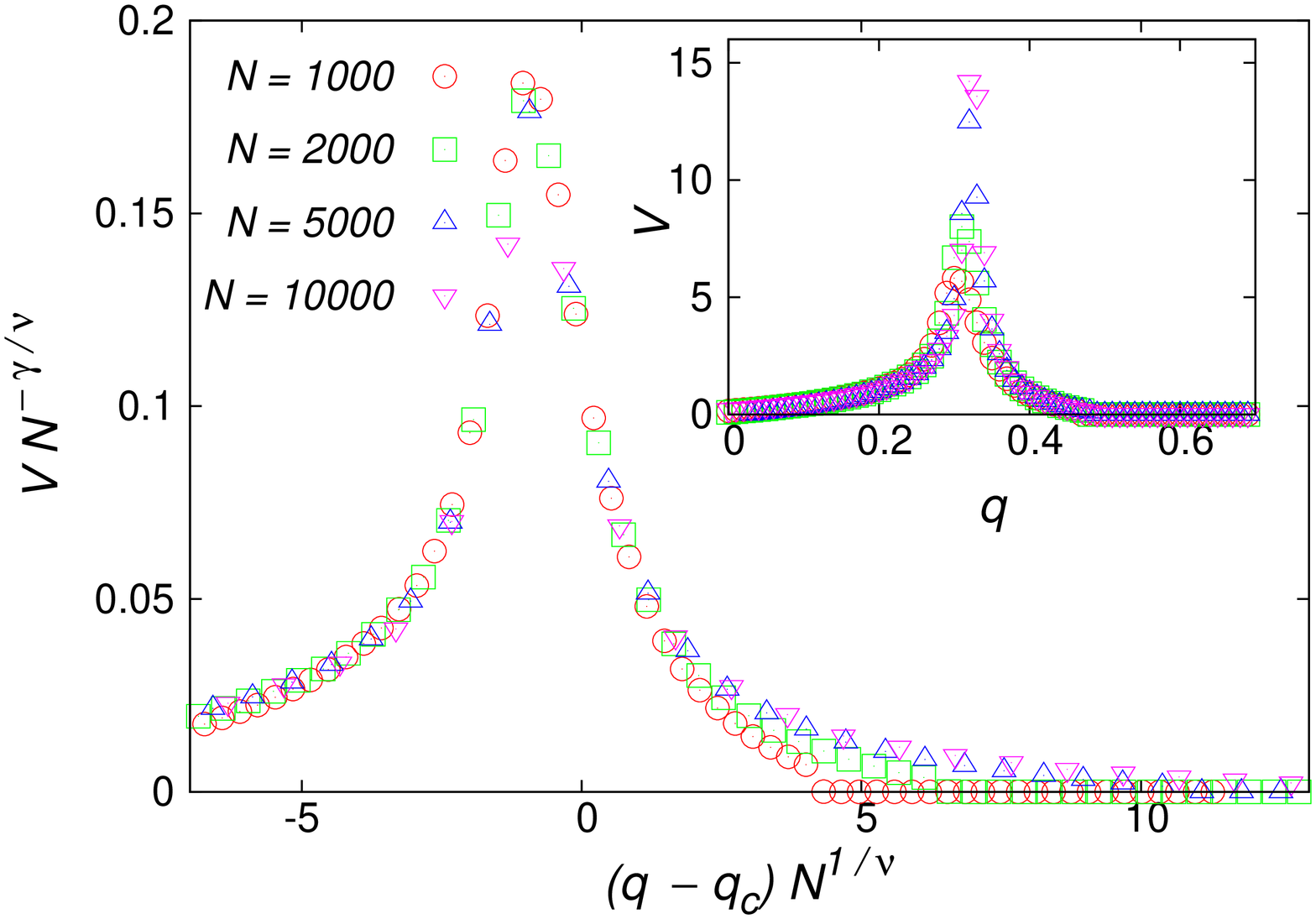}
\hspace{0.5cm}
\includegraphics[scale=0.25,angle=-90]{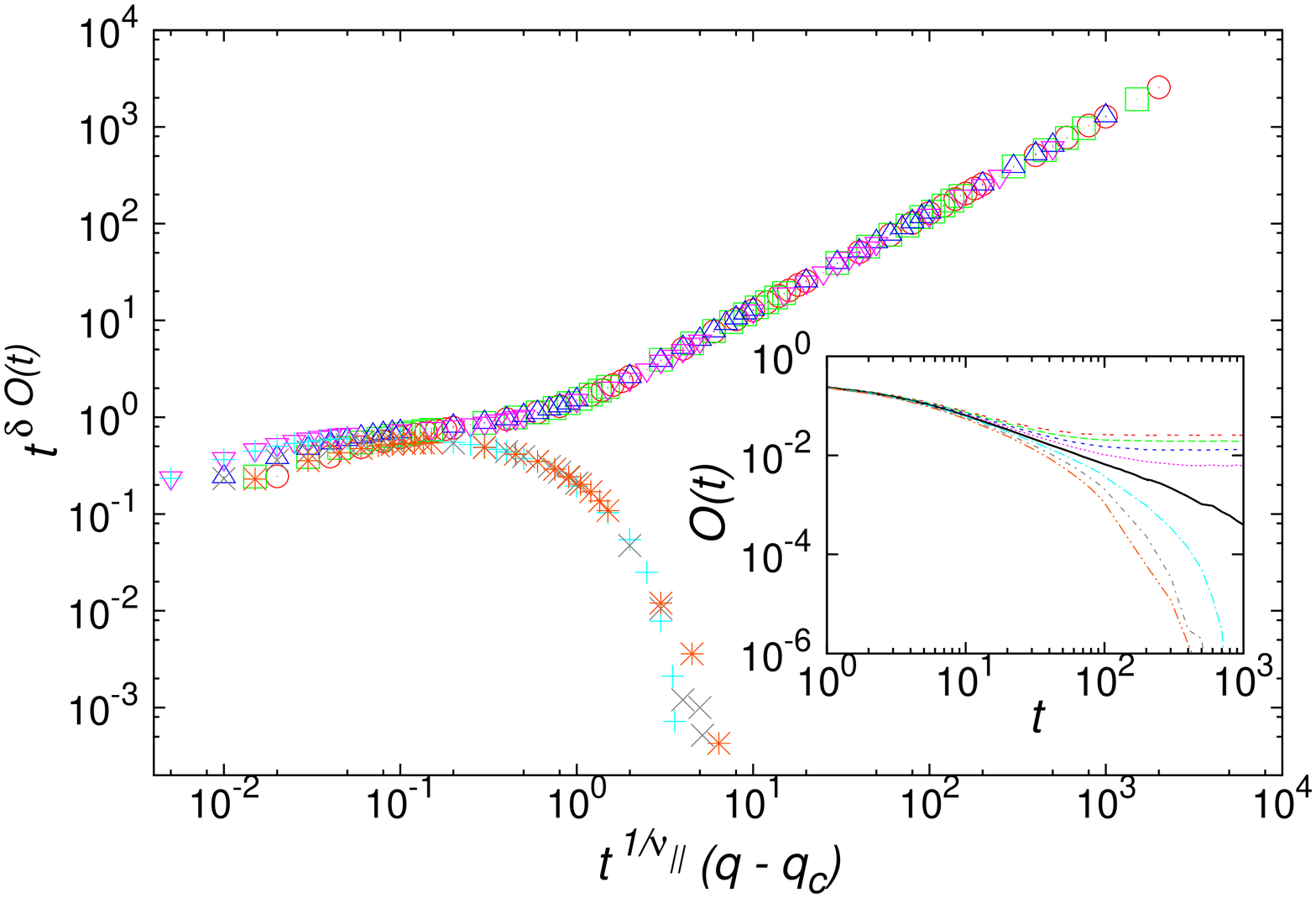}
\end{center}
\caption{(Color online) Scaling plot of some quantities of interest for $p=0.1$. (a) Susceptibility $V$ near a ferro-paramagnetic transition for typical system sizes $N$. The best data collapse was obtained for $\gamma\approx 1$, $1/\nu\approx 1/2$ and $q_{c}\approx 0.33$, considering Eqs. (\ref{eq19}) and (\ref{eq20}). (b) Order parameter $O$ for $N=10^{5}$ for distinct values of $q$ near a para-absorbing transition. The best data collapse was obtained for $\delta\approx 1$, $1/\nu_{||}\approx 1$ and $q_{c}\approx 0.5$, considering Eqs. (\ref{eq22}) and (\ref{eq23}). All data are averaged over $100$ independent simulations.}
\label{fig3}
\end{figure}

Finally, we performed simulations in order to obtain estimates of the other critical exponents, since our analytical results give us only the critical points and the exponent $\beta$. For the ferro-paramagnetic transition, we considered the usual finite-size scaling (FSS) relations,
\begin{eqnarray} \label{eq18}
O(N) & \sim & N^{-\beta/\nu} ~, \\  \label{eq19}
V(N) & \sim & N^{\gamma/\nu} ~, \\   \label{eq20}
q-q_{c} & \sim & N^{-1/\nu} ~,
\end{eqnarray}
\noindent
that are validy in the vicinity of the transition. In addition, for the transitions to the absorbing state (ferro-absorbing and para-absorbing), we considered the dynamic FSS relations \cite{dickman,hinrichsen}
\begin{eqnarray} \label{eq22}
O(t) & \sim & t^{-\delta} ~,\\ \label{eq23}
q-q_{c} & \sim & t^{-1/\nu_{||}} ~.
\end{eqnarray}
In Fig. \ref{fig3} we exhibit as an example results for $p=0.1$. In pannel (a) we show the data collapse for the susceptibility $V$, near the ferro-paramagnetic transition, where we obtained the usual mean-field Ising exponents $\gamma \approx 1$ and $\nu\approx 2$, that are the standard exponents of KEOM \cite{lccc,biswas2,biswas3,nuno_celia_victor}. In Fig. \ref{fig3} (b) we show the time relaxation of the order parameter (in the inset) in the vicinity of the para-absorbing transition, as well as the data collapse (main figure), for values near the transition point $q_{c}=1/2$. Our estimates for the critical exponents $\delta$ and $\nu_{||}$ are in agreement with the values for the directed percolation (and contact process), $\delta\approx 1$ and $\nu_{||}\approx 1$ \cite{dickman,hinrichsen}.

\begin{figure}[t]
\begin{center}
\vspace{6mm}
\includegraphics[scale=0.33,angle=-90]{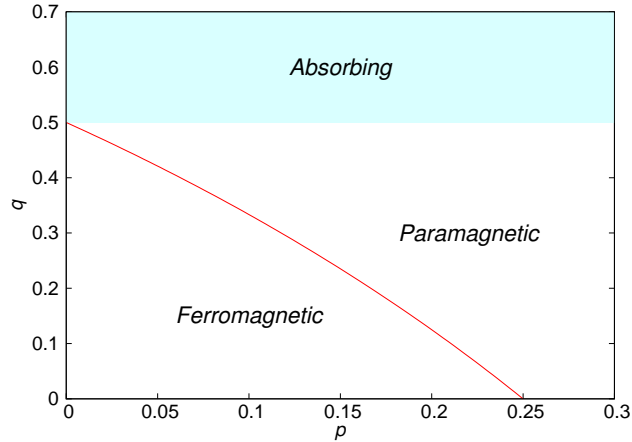}
\end{center}
\caption{(Color online) Phase diagram of the model in the plane $q$ versus $p$. The (blue) region represents the absorbing phase, and the full (red) curve represents the ferro-paramagnetic frontier, Eq. (\ref{eq7}). Notice that for $p=0$ the system presents an active-absorbing phase transition.} 
\label{fig4}
\end{figure}

To summarize, in Fig. \ref{fig4} we exhibit the phase diagram of the model in the plane $q$ versus $p$. The line separating the ferromagnetic and paramagnetic phases are given by Eq. (\ref{eq7}), valid in the region $p < 1/4$ and $q < 1/2$. The region for $q>\bar{q}_{c}=1/2$ represents the absorbing phase, for all values of $p$, and we also see the ferro-absorbing transition in the $p=0$ axis.

\section{Final remarks}

\qquad In this work we have studied how the inclusion of negative pairwise interactions (disorder) and indecision (noise) affect the critical behavior of a kinetic exchange opinion model. The agents can be in one of three possible states (opinions), represented by discrete variables $o=+1, -1$ and $0$. The topology of the society is a fully-connected network. The disorder is ruled by a parameter $p$, representing the fraction of negative interactions, and the noise is controlled by a parameter $q$, representing the probability of a spontaneous change to the neutral state. We analyzed the critical behavior of the system in the stationary states.

The consensus regarding one of the extreme opinions $+1$ or $-1$, a macroscopic collective behavior representing a total agreement, is only reached when the two ingredients, disorder and noise, are completely absent ($p=q=0$). For suitable values of $p$ and $q$, the system is in an ordered ferromagnetic phase, where one of the two extreme opinions $+1$ or $-1$ are shared by the majority of the population. This state mimics the situation where one of the two topics under debate is winner, and presents an order parameter $0<O<1$. We also found critical values $q_{c}(p)$ that separate the ordered ferromagnetic phase from a disordered paramagnetic one. In this last phase, the fractions of extreme opinions are equal and we have a null order parameter, $O=0$. In this case, the debate does not present a winner side. The critical exponent associated with the order parameter was found analytically to be $\beta=1/2$, and we estimated through Monte Carlo simulations that the other exponents are $\gamma\approx 1$ and $\nu\approx 2$, typical mean-field exponents belonging to the Ising model universality class. In this case, our numerical results show the typical finite-size effects for the order parameter when the system undergoes the order-disorder transition.

On the other hand, for $p=0$, i.e., in the absence of negative interactions, the system is governed only by the noise $q$. In this case, our analytical and numerical calculations showed that there is a critical point $q_{c}(p=0)=1/2$ above which the system is in an absorbing state. In this state, all agents become neutral, and the dynamics does not evolve anymore, characterizing a typical noise-induced absorbing phase transition. In this case, the order parameter is identically null, and it does not present the usual finite-size effects of order-disorder transitions (the ``tails'' of the order parameter). In this case, we considered the dynamic FSS relations for transitions to absorbing states. Our analytical results predicted an exponent $\beta=1$ for the order parameter in the vicinity of the transition, and numerically we found estimates for other exponents, namely $\delta\approx 1$ and $\nu_{||}\approx 1$. These exponents are the typical ones for the mean-field directed percolation, a prototype of a nonequilibrium phase transition to an absorbing state \cite{absorbing_book,dickman,hinrichsen}.

Finally, considering the case $p \neq 0$ and $q \geq q_c(p)$, when the system is in a paramagnetic state, we observed another transition. In this case, for $q>1/2$ the system is always in the absorbing state, $\forall p$. This phase is the same observed for the case $p=0$, but now the system goes from a paramagnetic phase with null order parameter to an absorbing phase also with $O=0$. In this case, in order to calculate the critical exponent $\beta$, we defined another order parameter based on the stationary fraction $f_{0}$ of neutral agents, namely $\bar{O} = (1-f_{0})/2$. In this case, we have $\bar{O}\neq 0$ in the paramagnetic phase, where the extreme opinions $+1$ and $-1$ coexist with the neutral state $0$, and we have $\bar{O}=0$ in the absorbing phase, where $f_{0}=1$. In this case, we also observed an exponent $\beta=1$, and the paramagnetic-absorbing transition also belongs to the directed percolation universality class. For the best of our knowledge, it is the first time that this para-absorbing transition appears in a KEOM.

All our analysis was done considering the new parameter $q$ as the control parameter, but the exponents do not depend on that, the another parameter $p$ can be considered as well, and the same critical exponents are found. The universality of the exponents is expected due to the mean-field character of the model. As extensions of this work, it can be considered the inclusion of local or global mass media effects \cite{nuno_jstat_2013,nuno2010}, as well as the inclusion of a neighborhood (lattice, network).

\section*{Acknowledgments}

The authors acknowledge financial support from the Brazilian funding agency CNPq.

\end{document}